\documentclass[pdf, twocolumn,superscriptaddress,floatfix,showpacs,amsmath,longbibliography,nofootinbib,amssymb]{revtex4-1}
\usepackage[colorlinks=true,citecolor=blue,linkcolor=blue]{hyperref}
\usepackage[usenames]{color}
\usepackage{subfigure}
\usepackage{dcolumn}
\usepackage{mathrsfs}% Align table columns on decimal point
\usepackage{bm}% bold math
\usepackage{amsmath,amssymb,epsfig,float,graphics}
\usepackage{appendix}
\newcommand{\bk}{\mathbf{k}}

\DeclareMathOperator{\arctanh}{arctanh}

\newcommand{\colb}[1]{\textcolor{blue}{#1}}

\begin{document}
\baselineskip=0.45 cm

\title{Geometric phase assisted detection of Lorentz-invariance violation from modified dispersion at high energies}

\author{Yihao Wu}
\affiliation{School of Physics, Hangzhou Normal University, Hangzhou, Zhejiang 311121, China}

\author{Zehua Tian}
\email{tzh@hznu.edu.cn}
\affiliation{School of Physics, Hangzhou Normal University, Hangzhou, Zhejiang 311121, China}

\begin{abstract}
Many theories of quantum gravity propose Lorentz-violating dispersion relations of the form $\omega_{|\bk|}=|\bk|f(|\bk|/M_\star)$, which approximately 
recover to the Lorentz invariance, $\omega_{|\bk|}\approx|\bk|$, at the energy scales much below $M_\star$. However, usually such a scale is assumed to be near the Planck scale, thus the feature of the Lorentz-violating theory is weak and its experimental test becomes extremely challenging. Since the geometric phase (GP) is of accumulative and sensitive nature to weak effects, here we explore the GP acquired by an inertial atomic detector that is coupled to a quantum field with this kind of Lorentz-violating dispersion. We show that for the Lorentz-violating field theory case the GP depends on the velocity of the detector, which is quite different from the Lorentz symmetry case where the GP is independent of the detector's velocity. In particular, we show that the GP may present a drastic low-energy Lorentz violation for any $f$ that dips below unity somewhere. We apply our analysis to detecting the polymer quantization motivated by loop quantum gravity, and show the detector acquires an experimentally detectable GP with the assist of detector's velocity that below current ion collider rapidities. Furthermore, the accumulative nature of GP might facilitate the relevant detection significantly.

%In order to probe Lorentz-violation effects experimentally we provide a scheme by geometric phase of an Unruh-DeWitt detector in massless sclar quantum field. We show %that the detector's geometric phase will  change if Lorentz-violation happened in sclar field. After qualitative analysing, we studied polymer quantization as an application. %By comparing Lorentz-violated case with invariant, we find that detector's velocity and initial state can assist in probing. Once detector's velocity pass the critial value, the %larger theory's explicit energy scale is, the easier we can probe it. Numerical estimation points that detecting such Lorentz-violation experimentally is viable. We also find %there exists a special velocity relates to energy scale passing which detector's geometric phase undergo a specific process that can be utilized in probing Lorentz-violation %at lower energy.

\end{abstract}

%\pacs{85.25.Dq, 84.40.Az, 04.80.Cc, 04.62.+v, 98.80.Cq}
\baselineskip=0.45 cm
\maketitle
\newpage

\section{Introduction} 
Lorentz invariance  is thought as one of the fundamental symmetries of relativity, however, many quantum gravity models, including string theory, warped brane worlds, loop quantum gravity, and so on, propose that Lorentz invariance may be broken at sufficiently high energies \cite{Camelia, Mattingly}. As one of the characters of quantum gravity, any discovery of Lorentz violation thus would be an important signal of beyond standard model physics, or constraining local Lorentz violations observationally may provide a potential constraint on quantum theories of gravity. Recently, there has been widespread interest trying to test Lorentz violation in diverse physical areas, spanning atomic physics, nuclear physics, high-energy physics, relativity, and astrophysics, see Refs. \cite{Mattingly, Colladay1, Colladay2, Jaffino, GW, Kostelecky}, and references therein.

In the Lorentz-violating theory, usually there might contain an explicit energy scale $M_\star$, such as the Planck energy, that characterizes the violation. When the energies much below such energy scale, the Lorentz invariance could be secured approximately. This energy scale sets an extremely high bar for testing the Lorentz violation, since the Planck energy, about $10^{19}~\mathrm{GeV}$, is much larger than any currently experimentally accessible energy scales, e.g., $10^{11}~\mathrm{GeV}$ for the trans-Greisen-Zatsepin-Kuzmin cosmic rays, which is the highest known energy of particles \cite{Mattingly}. Any experiments, in this respect, seems to be impossible to test the Lorentz violation at the Planck scale. Fortunately, in quantum field theory, strong Planck scale Lorentz violation could yield a small amount of violation at much lower energies, and thus it is possible for the effective low-energy theory to reveal unexpected imprints of the theory's high-energy structure \cite{Mattingly, Collins, Polchinski}. 

Recently, the response of an Unruh-DeWitt (UDW) detector \cite{Unruh, DeWitt, Birrell, Takagi, Crispino, Hu}, coupled to a class of Lorentz-violating quantum field theory in four-dimensional Minkowski spacetime, has been proposed to capture such unexpected low-energy imprints \cite{Husain}. It is found that an inertial UDW detector could present a drastic low-energy Lorentz violation when its velocity exceeds a critical one. One paradigm is that the polymer quantum field theories \cite{Ashtekar, Abhay, Viqar, Mortuza, Jasel} motivated by loop quantum gravity program \cite{Rovelli1, Rovelli2, Thiemann} predict a sharp transition behavior of low-energy UDW detector \cite{Husain, Nirmalya1, Stargen, Jorma, Golam, Nirmalya2}. Furthermore, as the proof of principle, analogue low-energy Lorentz violation from the modified dispersion at high energies, as suggested by quantum theories of gravity, has been proposed to be reproduced in dipolar Bose-Einstein condensates \cite{Tian1, Tian2}. Therefore, this feature of UDW detector may be expected to have highly potential application for the falsifiability of theoretical proposals, e.g., quantum gravity \cite{Husain, Jorma, Belenchia}. 

The previous investigation has been mainly focused on the qualitatively distinct character of the UDW detector caused by the Lorentz violation in quantum field theory. Whether the relevant signal of the UDW detector's transition rates could be strong enough to be observable in the possible experiment still remains elusive.
Besides, in order to excite response function of the UDW detector usually energy transfer is required, thus using such response function to measure the outside fields is limited, especial for the ultraweak fields where excitation events are rare. What is more, since the response function is indifferent to noise and signal, the detector is insensitive to fields in a noisy environments \cite{James, Dipankar}. Detecting the low-energy Lorentz violation from the modified dispersion at high energies through the response function seems to be restricted in practice. To make a progress on such detection, new detection methods and measurable physical quantities which are more sensitive to ultraweak fields or fields in noisy environments are needed to be explored and developed.

In this paper, we study the geometric phase  (GP) \cite{Berry} acquired by an inertially moving atomic detector that is coupled to a quantum field with the Lorentz-violating dispersion, $\omega_{|\bk|}=|\bk|f(|\bk|/M_\star)$.  This phase for a closed quantum system depends only on the geometry of the parameter space \cite{Berry}, while it depends not only on the unitary evolution, but also on transition and decoherence rates if the system interacts with an environment \cite{Carollo}. From an experimental point of view, this phase is much more sensitive than the transition rate and could also be amplified through accumulation. There thus has been fruitfully applied to the detection and understanding of ultraweak field effects to bring down the scale experimentally required \cite{James, Dipankar, Eduardo1, HU, Antonio, Tian3, Jun, Partha, Arya, Partha, Debasish}. A typical example is that with the use of GP, the required acceleration to detect the Unruh effect in quantum field theory can be lowered to the order of $10^7 \mathrm{m/s^2}$, which is much lower than the previous acceleration requirement by many orders of magnitude \cite{Eduardo1, Arya}.
Here, the GP has been proposed to detect the ultraweak low-energy imprints of the Lorentz violation from modified dispersion at high energies. We find that: 1) in the presence of Lorentz-violating field theory, the acquired GP depends on the velocity of the detector, while it is independent of the detector's velocity for the Lorentz-invariant case; 2) the GP may present a drastic low-energy Lorentz violation for any $f$ that dips below unity somewhere; 3) when detecting the polymer quantization motivated by loop quantum gravity, the detector acquires an experimentally detectable GP with the assist of detector velocity below current ion collider rapidities; 4) the accumulative nature of the GP facilitates the detection of weak effects such as the polymer quantization. Therefore, our studies may provide a new method that susceptible to the Lorentz violation from the modified dispersion at high energies.

This paper is constructed as follows. In Sec. \ref{section2}, we introduce our model consisting of a two-level atom (or detector) and  the coupled massless scalar field with Lorentz-violating dispersion relations of the form $\omega_{|\bk|}=|\bk|f(|\bk|/M_\star)$. We also calculate the GP acquired by the detector that is induced by the Lorentz violation. In Sec.  \ref{section3}, we apply our scheme to the detection of polymer quantization theory that is motivated by loop quantum gravity. Finally, we conclude and discuss our results in Sec. \ref{section4}. We use the natural units $c=\hbar=1$ in the paper.

\section{Detector model and geometric phase} \label{section2}
In this section we will introduce our model consisting of a two-level atom (as the detector) and a massless scalar field with Lorentz-violating dispersion.
We also derive the dynamics of the detector moving with a constant velocity and calculate the GP the detector acquires.  
Our discussion is confined in the $(1+3)$-dimensional Minkowski spacetime with the metric, ${ds}^{2}={dt}^{2}-dx^{2}-dy^2-dz^{2}$.

\subsection{Lorentz violation in quantum field theory}\label{subsection1}
The Lorentz-violating quantum field theories we consider have a specific form of dispersion relation, which reads 
\begin{eqnarray}\label{Omega}
\omega_{|\bk|}=|\bk|f(|\bk|/M_\star).
\end{eqnarray}
Here $\omega_{|\bk|}$ is the corresponding energy with the spatial momentum $\bk$. The positive constant $M_\star$ determines the energy scale of Lorentz violation, and 
$f$ is a smooth positive function on the positive real line. Define a dimensionless quantity, $g=|\bk|/M_\star$, the function $f(g)$ has the limit, $\lim_{g\rightarrow0_+}f(g)\rightarrow1$, and correspondingly the dispersion relation approaches to the Lorentz-invariant case for $g=|\bk|/M_\star\ll1$, i.e., $\omega_{|\bk|}\approx|\bk|$.

%Quantum gravity theories usually provide Lorentz-violated dispersion relation with phenomenological form $\omega_{|\textbf{k}|}=|\textbf{k}|f(|\textbf{k}|/M_\star)$, 
%there $\omega_{|\textbf{k}|}$ is energy, $\textbf{k}$ is $3$-momentum, $M_\star$ is a scalar factor that owns the dimension of energy, which maybe compared with Planck %energy, $f(g)$ is a successive dimensionless function own the trait $f(g)\to1$ when $g\to0$ so that guarantee the theroy recover to Lorentz invariance as 
%$|\textbf{k}|/M_\star\to0$.

We assume that a real scalar field $\Phi$ can be decomposed into spatial Fourier modes such that a mode with spatial momentum $\bk\neq0$ is a harmonic oscillator with 
the angular frequency $\omega_{|\bk|}$ shown in \eqref{Omega}. In this case, its quantization reads 
\begin{eqnarray}\label{Decomposition}
\Phi(\textbf{x}) =\int{d^3\textbf{k}\,\rho_{|\textbf{k}|}(a^\dagger_{\textbf{k}}e^{-i\textbf{k}\cdot\textbf{x}}+a_{\textbf{k}}e^{i\textbf{k}\cdot\textbf{x}})},
\end{eqnarray}
where $\textbf{x}=(x, y, z)$, $\rho_{|\textbf{k}|}=d(|\textbf{k}|/M_\star)/\sqrt{(2\pi)^3|\textbf{k}|}$ is the density-of-states weight factor, which is a smooth complex-valued function 
on the positive real line. Note that if $f(g)=1$ and $|d(g)|=1/\sqrt{2}$, the field $\Phi$ describes the usual massless scalar field.

%Because of the Lorentz-violation of messless scalar field, the quantized field operator should be modified to 
%\begin{eqnarray}
%\Phi[\textbf{x}] =\int{d^3\textbf{k}\,\rho_{|\textbf{k}|}(a^\dagger_{\textbf{k}}e^{-i\textbf{k}\cdot\textbf{x}}+a_{\textbf{k}}e^{i\textbf{k}\cdot\textbf{x}})},
%\end{eqnarray}
%where $\rho_{|\textbf{k}|}=d(|\textbf{k}|/M_\star)/\sqrt{(2\pi)^3|\textbf{k}|}$ is state denisty, here $d(g)$ satisfy $d(g)\to1/\sqrt2$ if $g\to0$.

\subsection{Dynamics of two-level atom}\label{subsection2}
The detector we consider is assumed to be a two-level atom, and correspondingly its excited and ground states are respectively given by $|e\rangle$ and $|g\rangle$ with the energy-level spacing $\omega_0$. In order to capture the properties of Lorentz-violation in quantum field theory, we assume that the detector is coupled to the 
quantum field $\Phi$ shown in \eqref{Decomposition}. The Hamiltonian of the whole system then reads 
\begin{eqnarray}
H=H_{0}+H_\Phi+H_I.
\end{eqnarray}
Here $H_{0}$ is the detector's Hamiltonian, written as 
\begin{eqnarray}
H_{0}=\frac{1}{2}\omega_0\sigma_z,
\end{eqnarray}
with $\sigma_z$ being the Pauli matrix in $z$-component. $H_\Phi$ is field's Hamiltonian, and $H_I$ denotes the interaction Hamiltonian between the detector and field. 
Specifically, 
\begin{eqnarray}\label{IH}
 H_I=\mu\sigma_x\Phi(\textbf{x}),
\end{eqnarray}
where $\mu$ is the coupling constant that usually is assumed to be small,  and $\sigma_x=\sigma_++\sigma_-$ is the Pauli matrix in $x$-component, with $\sigma_+$$(\sigma_-)$ being the atomic rasing (lowering) operator.

At the beginning, the initial state of the detector and field is assumed to be $\rho_\text{tot}(0)=\rho(0)\otimes|0\rangle\langle0|$, where $\rho(0)=|\Psi(0)\rangle\langle\Psi(0)|$, with $|\Psi(0)\rangle=\cos\frac{\theta}{2}|g\rangle+\sin\frac{\theta}{2}|e\rangle$, denotes the detector's initial state, and $|0\rangle$ is the vacuum of the field.
The dynamics of the the whole system satisfies the Liouville equation, which in the interaction picture reads 
\begin{eqnarray}\label{EE}
\frac{\partial{\rho_\text{tot}(\tau)}}{\partial\tau}=-i[H_I(\tau),\rho_\text{tot}(\tau)].
\end{eqnarray}
Here $\tau$ denotes the proper time of the detector. It is needed to note that in the interaction picture $H_I$ in \eqref{IH} actually has been rotated to 
\begin{eqnarray}
H_I(\tau)=\mu(\sigma_+e^{i\omega_0\tau}+\sigma_-e^{-i\omega_0\tau})\Phi(t(\tau),\textbf{x}(\tau)),
\end{eqnarray}
shown in \eqref{EE}. Since we are interested in the dynamics of the detector, usually the degree of freedom of the field has to be traced over.
Besides, in the limit of  weak coupling, the Born approximation and Markov approximation \cite{Breuer} can be used to derive the master 
equation of the detector. After these processes, the reduced density matrix, $\rho(\tau)$, that describes the evolving state of the detector, finally satisfies 
the dynamics in the Kossakowski-Lindblad form \cite{Gorini, Lindblad, Breuer},
\iffalse
% Formally integrate the (7) with substitute back it and take trace toward enviornment, consider Born approximation \cite{19.1} and Markov approximation \cite{19.1}, we have
%\begin{eqnarray}
	%\frac{\partial\rho(\tau)}{\partial\tau} =-\int_{0}^{\tau} ds Tr_E[H_I(\tau),[H_I(\tau-s),\rho(\tau)\otimes\rho_E]],
%\end{eqnarray}
%Note that our research object is opening system and the correlation time between system and enviornment is very short, so the upper limit of integral can be infinity
%\begin{eqnarray}
	%\frac{\partial\rho(\tau)}{\partial\tau} =-\int_{0}^{\infty} ds Tr_E[H_I(\tau),[H_I(\tau-s),\rho(\tau)\otimes\rho_E]],
%\end{eqnarray}
%By subsitituting interaction Hamiltonian Eq.(5) into (9),take rotating wave approximation and after a serise of cacultion, return to Schrodinger picture we derive the master %equation for the atom
\fi
\begin{eqnarray}\label{LO}
\nonumber 
\frac{\partial\rho(\tau)}{\partial\tau} =-i[H_{\text{eff}},\rho(\tau)]+\sum_{j=1}^{3} (2L_j\rho L_j^\dagger-L_j^\dagger L_j\rho-\rho L_j^\dagger L_j).
\\
\end{eqnarray}
Here the effective Hamiltonian $H_{\text{eff}}=H_0+H_{\text{shift}}$ with the Lamb shift \cite{Breuer, Lamb}, $H_\text{shift}=\frac{1}{2}\mu^2\mathrm{Im}(\Gamma_++\Gamma_-)\sigma_z$, where $\Gamma_\pm$ will be defined below. However, usually the Lamb shift is much smaller than the energy-level spacing of the atom, $\omega_0$, which thus can be ignored.
Furthermore, the operators in \eqref{LO} have been defined as
\begin{eqnarray}
 L_1=\sqrt{\frac{\gamma_+}{2}}\sigma _+,~~~~L_2=\sqrt{\frac{\gamma_-}{2}}\sigma _-,~~~L_3=\sqrt{\frac{\gamma_z}{2}}\sigma_z,
\end{eqnarray}
with 
\begin{eqnarray}\label{gamma}
\nonumber
\gamma_\pm&=&2\mu^2\mathrm{Re}\Gamma_\pm=\mu ^2\int_{-\infty }^{+\infty} e^{\mp i\omega_0\Delta\tau}G^+(\Delta\tau-i\epsilon)d\Delta\tau,
\\ 
\gamma_z&=&0.
\end{eqnarray}
Note that $\Delta\tau=\tau-\tau^\prime$ has been defined, and  $G^+(\tau-\tau^\prime)=\langle 0|\Phi(t(\tau),\textbf{x}(\tau))\Phi(t'(\tau^\prime),\textbf{x}'(\tau'))|0 \rangle$ is field Wightman function along the worldline of the detector $(t(\tau),\textbf{x}(\tau))$. Actually, $\gamma_\pm$ are the Fourier transform of the field Wightman function, denoting the detector's transition rates.

For convenience, we can write the density matrix $\rho(\tau)$ in terms of the Pauli matrices,
\begin{eqnarray}\label{TS}
\rho(\tau)=\frac{1}{2}\bigg(\mathbf{I}+\sum^3_{i=1}\omega_i(\tau)\sigma_i\bigg),
\end{eqnarray}
where $\mathbf{I}$ is a $2\times2$ identity matrix, and $\sigma_i=(\sigma_x, \sigma_y, \sigma_z)$.  Substituting \eqref{TS} into Eq. \eqref{LO} and using the initial state 
assumed before, we can finally obtain the time-dependent parameters of the reduced density matrix,  
\begin{eqnarray}\label{ISP}
	\nonumber
	\omega_1(\tau)=\text{sin}\theta \text{cos}(\omega_0 \tau)e^{-\frac{1}{2} (\gamma _++\gamma _-)\tau},
\end{eqnarray}
\begin{eqnarray}
	\omega_2(\tau)=\text{sin}\theta \text{sin}(\omega_0 \tau)e^{-\frac{1}{2} (\gamma _++\gamma _-)\tau},
\end{eqnarray}
\begin{eqnarray}
	\nonumber
	\omega_3(\tau)=-\text{cos}\theta e^{-(\gamma _++\gamma _-)\tau}+\frac{\gamma _+-\gamma _-}{\gamma _++\gamma _-}[1-e^{-(\gamma _++\gamma _-)\tau}].
\end{eqnarray}
Note that here the field properties are contained in $\gamma_\pm$, and thus could be encoded into the state of the detector, which thus can be read out finally.

\subsection{Geometric phase}
When taken around a closed trajectory in the parameter space, in addition to the dynamical phase, the state of a quantum system could acquire an additional phase factor.
Since this phase depends only on the geometry of the parameter space, which is called as the GP \cite{Berry}. The GP for a mixed state undergoing a non-unitary evolution 
is \cite{Tong}
\begin{eqnarray}\label{GPD}
\nonumber
\Omega &=&\mathrm{arg}\bigg(\sum_{k=1}^{N}\sqrt{\lambda_k(0)\lambda_k(T)}\langle\phi_k(0)|\phi_k(T)\rangle 
\\ 
&&\times\,e^{-\int_{0}^{T}\langle\phi_k(\tau)|\dot{\phi}_k(\tau)\rangle d\tau}\bigg),
\end{eqnarray}
where $T$ is the evolution time of the detector, $\lambda_k$ and $|\phi_k(\tau)\rangle$ are respectively the eigenvalues and corresponding eigenvectors of the reduced 
density matrix $\rho(\tau)$ shown above. $|\dot{\phi}_k(\tau)\rangle$ is the derivative with respect to the detector's proper time $\tau$. 
In order to calculate the GP of the evolving state in Eq. \eqref{TS}, one has to first obtain its eigenvalues, which is found to be 
\begin{eqnarray}
\lambda_\pm(\tau)=\frac{1}{2} (1\pm \eta ),
\end{eqnarray}
with $\eta =\sqrt{e^{-(\gamma _++\gamma _-)\tau}\text{sin}^2\theta+\omega _3^2(\tau)} $. 
Note that $\omega _3(0)=-\text{cos}\theta$, thus $\eta(0)=1$, $\lambda_-(0)=0$, which means it has no contribution to the GP. 
Therefore, we just need to consider the eigenvector corresponding to $\lambda_+(\tau)$, which reads 
\begin{eqnarray}
|\phi_+(\tau)\rangle=\text{sin}\frac{\theta_\tau}{2}|e\rangle+\text{cos} \frac{\theta_\tau}{2}e^{i\omega_0\tau}|g\rangle,
\end{eqnarray}
with $\theta_{\tau}$ being
\begin{eqnarray}
\text{tan}\frac{\theta_\tau}{2}=\sqrt{\frac{\eta +\omega_3}{\eta -\omega_3} }.
\end{eqnarray}
According to the definition shown in Eq. \eqref{GPD}, the GP for a time interval $T$ of evolution is 
\begin{eqnarray}\label{GGP}
	\Omega =-\frac{\omega_0}{2}\int_{0}^{T} d\tau\bigg(1+\frac{\text{cos}\theta+\Delta-\Delta e^{(\gamma_++\gamma_-)\tau} }{\sqrt{R^2+e^{(\gamma_++\gamma_-)\tau}\text{sin}^2\theta}}\bigg),
\end{eqnarray}
where $\Delta=(\gamma_+-\gamma_-)/(\gamma_++\gamma_-)$, and $R=\text{cos}\theta+\Delta[1-e^{(\gamma_++\gamma_-)\tau}]$. Note that 
the GP in \eqref{GGP} is general. Different spacetime background and the detector's trajectories may cuase different evolution of the 
detector, and thus induce different GP. Which, in turn, from the correction of the GP one can characterize the properties of 
spacetime background and the detector's state of relativistic motion. In the following, we will calculate the GP for a detector which is coupled 
to a Lorentz-violating field and moves with a constant velocity in the $(1 + 3)$-dimensional Minkowski spacetime.

\subsection{Geometric phase for the Lorentz-violating field theories} \label{subsectionA}
How the Lorentz violation affects the GP of a two-level atom (detector)?  To address this issue, we will consider the detector's dynamics introduced in section \ref{subsection2}, while the detector in this scenario interacts with the Lorentz-violating field described in section \ref{subsection1}.
Furthermore, we also assume that the detector moves with a constant velocity, the corresponding trajectory reads 
\begin{eqnarray}\label{trajectory}
	(t(\tau),\textbf{x}(\tau))=(\tau \text{cosh}\beta,0,0,\tau \text{sinh}\beta),
\end{eqnarray}
where $\beta$ is the rapidity with respect to the distinguished inertial frame.

Substituting the above trajectory into \eqref{gamma}, we can obtain 
\begin{eqnarray}\label{RF}
\nonumber	
\gamma_\pm&=&\frac{\mu^2M_\star}{2\pi\text{sinh}\beta}\int_{0}^{\infty}dg|d(g)|^2
\\
&&\times \text{H}(g\text{sinh}\beta-|(\pm\omega_0/M_\star)+gf(g)\text{cosh}\beta|),
\end{eqnarray}
where $\text{H}(x)$ is Heaviside function, and $g=|\textbf{k}|/M_\star$.
For convenience, we rewrite $\gamma_\pm=\mu^2M_\star F_\pm$ with
\begin{eqnarray}\label{RF2}
\nonumber	
F_\pm&=&\frac{1}{2\pi\text{sinh}\beta}\int_{0}^{\infty}dg|d(g)|^2
\\
&&\times \text{H}(g\text{sinh}\beta-|h+gf(g)\text{cosh}\beta|).
\end{eqnarray}
Here $h$ is defined as a dimensionless argument: for the excitation rate $\gamma_+$,  $h=\omega_0/M_\star$; for the deexcitation rate $\gamma_-$, $h=-\omega_0/M_\star$. Note that the scale parameter $M_\star$ enters \eqref{RF} only as the overall factor and via the combination $\omega_0/M_\star$.

For the usual massless scalar field, i.e., $f(g)=1$, and $|d(g)|=1/\sqrt{2}$ in \eqref{Decomposition}, we can find that $\gamma_+$ vanishes, which means that 
the detector does not become spontaneously excited. However, in this case, $\gamma_-=\mu^2\omega_0/2\pi$. It denotes the spontaneous decay rate of the detector and 
is independent of the velocity $\beta$. For the Lorentz-violating case, we can find that the crucial issue in \eqref{RF} is the behavior 
of the argument of the Heaviside function $\mathrm{H}$: Under what condition is the argument of $\mathrm{H}$ positive for at least some interval of $g$? 
If $f(g)\ge1$ for all $g$, we can find that for the $\gamma_+$ case, the argument of $\mathrm{H}$ always negative for all $h$. Thus the transition rate $\gamma_+$
vanishes, suggesting that the detector does not become spontaneously excited. However, the decay rate $\gamma_-$ in this case always exists and depends on the velocity of the detector as a result of the Lorentz violation.

An interesting case appears when $f(g)$ can dip somewhere below unity \cite{Husain, Tian1}, with $f_c=\inf\,f$, and $0<f_c<1$. In this scenario, we can find that there is a critical 
velocity $\beta_c=\arctanh(f_c)$, below which, i.e., $0<\beta<\beta_c$, $\gamma_+$ vanishes and thus the inertial detector moving with a constant velocity remains unexcited. The deexcitation rate is found to be $\gamma_-=\frac{\mu^2\omega_0}{2\pi}\{1+\cosh\beta[(1+2\cosh(2\beta))f^\prime(0)-2\text{Re}(d^\prime(0)/d(0))]h+\mathcal{O}(h^2)\}$ for the small $\omega_0$ and the velocity does not approach crucial one very much, which shows that it is rapidity dependent, and the Lorentz violation is suppressed at low energies by the factor $\omega_0/M_\star$. Obviously, when $M_\star\rightarrow\infty$ with fixed $\omega_0$, it reduces to the deexcitation rate of a detector coupled to the usual massless scalar field \cite{Birrell}. However, if the velocity of the detector exceeds the critical velocity, i.e., $\beta>\beta_c$, the argument of $\mathrm{H}$ in \eqref{RF2} always 
keeps position for $0<h<\sup_{g\ge0}g[\sinh\beta-f(g)\cosh\beta]$, and thus $\gamma_+$ does not vanish. It means in this case the inertial detector with arbitrarily small positive $h$ will be excited. This is quite different from the result of the uniformly moving detector that interacts with the usual massless scalar field, which never gets spontaneously excited as discussed above. Besides, we also note that the deexcitation rate $\gamma_-$ in this case does not vanish and depends on the detector's rapidities.

From the above discussions, we can find that in the presence of the Lorentz violation the transition rate of the inertially moving detector may depend on the detector's rapidity, which 
is different from the Lorentz invariant case, where the detector's transition rate is rapidity-independent. Particularly, 
if a quantum field with the dispersion in \eqref{Omega} satisfies the conditions: the smooth positive-valued function $f(g)\rightarrow1$ as $g\rightarrow0_+$, and $f_c=\inf\,f$ yielding $0<f_c<1$, and $|d(g)|\rightarrow1/\sqrt{2}$ as $x\rightarrow0_+$, $d(g)$ is nonvanishing everywhere, an inertial UDW detector with rapidity $\beta>\beta_c=\arctanh(f_c)$ in the preferred frame experiences spontaneous excitation at arbitrarily low $\omega_0$. However, this never occurs in the Lorentz-invariant quantum field theory. Furthermore, both the excitation and deexcitation rates are proportional to the Lorentz 
violation energy scale $M_\star$. Since the GP is sensitive to transition rates and from the above analysis the transition rates becomes significantly modified by the Lorentz violation and the detector's velocity, we expect the Lorentz-violating component of the GP to be correspondingly modified. Besides, the accumulative nature of the GP \cite{Eduardo1, Arya} could facilitate the detection of weak effects such as the Lorentz-violating modifications to the field.

%The crucial question is wheather Heaviside function is non-zero such that the detector owns non-zero transition rate: If the  sclar field is Lorentz-invariant, because the %spectrum we researching owns traits $f(g)\to1$ and $d(g)\to1/\sqrt2$ as $g\to0$, Heaviside function remainds zero towards $\gamma_+$, which means detector won't be %excited as we expected, and the transition rate of deexcitation $\gamma_-$ is a constant $\mu^2\frac{\omega_0}{2\pi}$ not affected by velocity $\beta$ for all energy gap $%\omega_0$. While the  sclar field is Lorentz-violated, the argument of Heaviside function will pass zero for some velocity $\beta$ to any energy gap $\omega_0$, meaning %detector is excited spontaneously and transition rates are influenced by detector's velocity.

For convenience, we rewrite the deexcitation rate of the detector coupled to the usual massless scalar quantum field as $\gamma_0=\frac{\mu^2\omega_0}{2\pi}$.
Then the transition rates for the Lorentz-violating case in Eqs. \eqref{RF} and \eqref{RF2} can be rewritten as 
\begin{eqnarray}\label{cpm}
\gamma_\pm=\mu^2M_\star F_\pm=\pm\gamma_0\frac{2\pi F_\pm}{h}=\gamma_0c_\pm,
\end{eqnarray}
where $c_\pm=\pm\frac{2\pi F_\pm}{h}$ are dimensionless functions which are determined by the detector's rapidity $\beta$ and 
the detector's energy-level spacing $|h|$. Since the parameter $\gamma_0/\omega_0=\mu^2/2\pi$ is small, we can Taylor expand the GP in \eqref{GGP}
to the first order of $\gamma_0/\omega_0$ \cite{Hu, James, Tian3,  Arya}. For the uniformly moving detector coupled to a quantum field with Lorentz-invariant dispersion, we can obtain the corresponding GP,
\begin{eqnarray}\label{InertialGP}
\Omega_I=-\varphi\text{cos}^2\frac{\theta}{2}+\varphi^2\frac{\gamma_0}{8\omega_0}\text{sin}^2\theta(\text{cos}\theta-2),
\end{eqnarray}
where $\varphi=\omega_0T$. The first term $\Omega_D=-\varphi\text{cos}^2\frac{\theta}{2}$ in \eqref{InertialGP} denotes the GP of an isolated detector, and second term is the correction, $\delta\Omega_I=\varphi^2\frac{\gamma_0}{8\omega_0}\text{sin}^2\theta(\text{cos}\theta-2)$, as a result of the interaction between detector and quantum field. Here the GP for Lorentz-invariant field theory case just depends on the accumulation time,
the initial state and energy-level spacing of the detector, and the coupling strength between the detector and field.

If an uniformly moving detector is coupled to a quantum field with a dispersion in \eqref{Omega}, we can obtain the corresponding GP of the detector, which 
is Taylor expanded to the first order of $\gamma_0/\omega_0$,
\begin{eqnarray}\label{LVGP}
\nonumber
\Omega_V&=&-\varphi\text{cos}^2\frac{\theta}{2}+\varphi^2\frac{\gamma_0}{8\omega_0}\text{sin}^2\theta\big[2(c_+-c_-)
\\
&&+\text{cos}\theta(c_++c_-)\big].
\end{eqnarray}
The first term denotes the GP of an isolated detector, $\Omega_D=-\varphi\text{cos}^2\frac{\theta}{2}$. The second term is the correction to 
the GP that arises from the interaction between detector and external quantum field, which is given by
\begin{eqnarray}\label{correction}
\delta\Omega_V=\varphi^2\frac{\gamma_0}{8\omega_0}\text{sin}^2\theta[2(c_+-c_-)+\text{cos}\theta(c_++c_-)].
\end{eqnarray}
This correction is quite different from that of the Lorenz-invariant case. Besides the accumulation time,
the initial state and energy-level spacing of the detector, and the coupling strength between the detector and field, it also depends on the 
rapidity of the detector, seen from \eqref{RF}. Note that the rapidity-dependent property of the GP results from the Lorentz violation.

From the above analysis, it is found that when a quantum field is of a Lorentz-violating dispersion, the detector that the field is coupled to could acquire 
a correction of GP which depends on the detector's rapidity. However, for the Lorentz invariant case, the corresponding GP correction is independent of the rapidity.
Therefore, whether the GP is rapidity dependent could be used as a criteria to  characterize the Lorentz violation. Due to this feature, 
the GP here can be used to probe the Lorentz violation with the assistance of the detector's rapidity. Besides, the GP could be accumulative, this nature may also 
facilitate the detection of weak effects. In what follows, we will apply the above analysis to a concrete example---to the detection of the polymer quantization motivated by loop quantum gravity.

%%%%%%%%%%%%%%%%%%%%%%%%%%%%%%%%%%%%%%%%%%%%%%%%%%%%%%%%%%%
\section{Geometric phase in polymer quantum field theory} \label{section3}
Polymer quantization \cite{Ashtekar, Halvorson} is used as a quantization method in loop quantum gravity \cite{Rovelli1, Rovelli2, Thiemann}, which is a background independent canonical quantization of General Relativity and a candidate theory of quantum gravity \cite{Ashtekar2}. This quantization method differs from the usual Schr\"odinger quantization in several important ways when applied to a mechanical system. Firstly, apart from \emph{Planck constant}, $\hbar$, it usually involves a new dimension-full parameter, e.g., 
the \emph{Planck length}, $L_p=\sqrt{\hbar G/c^3}$ ($G$ is Newton's gravitational constant, and $c$ is the speed of light in vacuum), in the context of quantum gravity. Secondly, the position and momentum operators can not be both defined simultaneously in polymer quantization, but rather only one of them. This is because the kinematical 
Hilbert space is nonseparable. Therefore, usually only one of the position and momentum is represented directly as an elementary operator in the kinematical Hilbert space, while the second one is represented as exponential of its classical counterpart. It is natural to expect the different set of results from the polymer quantization compared to
those from Schr\"odinger quantization. 

Polymer quantization has been applied to many fields, e.g., cosmology \cite{Martin, Abhay, Ma1, Sanjeev}, black hole physics \cite{Ma2, Zhang, Wu, Ma3}, scalar field \cite{Mortuza, Ashtekar, Stargen}, and so on. Here we will consider the specific implementation of a polymer quantized scalar field studied in Ref. \cite{Mortuza}, which 
has been fruitfully applied to the exploration of the transition rates of UDW detector with different worldlines, including inertial \cite{Husain, Nirmalya1, Jorma, Nirmalya2} and accelerated cases \cite{Stargen, Jorma, Golam, Golam2}. We shall explore the response of the GP of a two-level detector introduced above to the polymer quantization with the assist of the detector's rapidity.

\subsection{Transition rates in the polymer quantum field theory}

The Wightman function in polymer quantum field theory reads \cite{Mortuza}
\begin{eqnarray}\label{WF}
\nonumber
G^+(t, \textbf{x}; t^\prime, \textbf{x}^\prime)&=&\frac{1}{(2\pi)^3} \int d^3\textbf{k}e^{i\textbf{k}\cdot({\textbf{x}-\textbf{x}^\prime)}}
\\
&&\times \sum_{n=0}^{\infty } |c_{n}(|\textbf{k}|)|^2e^{-i\Delta E_{n}(|\textbf{k}|)(t-t^\prime)}.
\end{eqnarray}
Here, $\Delta E_n(|\bk|)=E_n(|\bk|)-E_0(|\bk|)$, with
\begin{eqnarray}
E_{2n}(|\textbf{k}|)/\omega&=&\frac{2g^2A_n(1/4g^2)+1}{4g},
\\     
E_{2n+1}(|\textbf{k}|)/\omega&=&\frac{2g^2B_{n+1}(1/4g^2)+1}{4g},
\\
c_{n}(|\textbf{k}|)&=&1/\sqrt{M_\star}\int_{0}^{2\pi}  \psi_n(i\partial_{u})\psi_0du,
\end{eqnarray}
and $\psi_n$ are given by
\begin{eqnarray}
\psi_{2n}(u)=\pi^{-1/2}\mathrm{ce}_{n}(1/4g^2,u),
\end{eqnarray}
\begin{eqnarray}
\psi_{2n+1}(u)=\pi^{-1/2}\mathrm{se}_{n+1}(1/4g^2,u),
\end{eqnarray}
satisfying the Mathieu equation, referring to Ref. \cite{Mortuza} for the details.
$B_{n}(x)$ and $A_{n}(x)$ are the Mathieu characteristic value functions, $\mathrm{ce}_{n}(x,q)$ and $\mathrm{se}_{n}(x,q)$ are respectively the elliptic cosine and sine functions.

\begin{figure}
\centering
\includegraphics[width=0.39\textwidth]{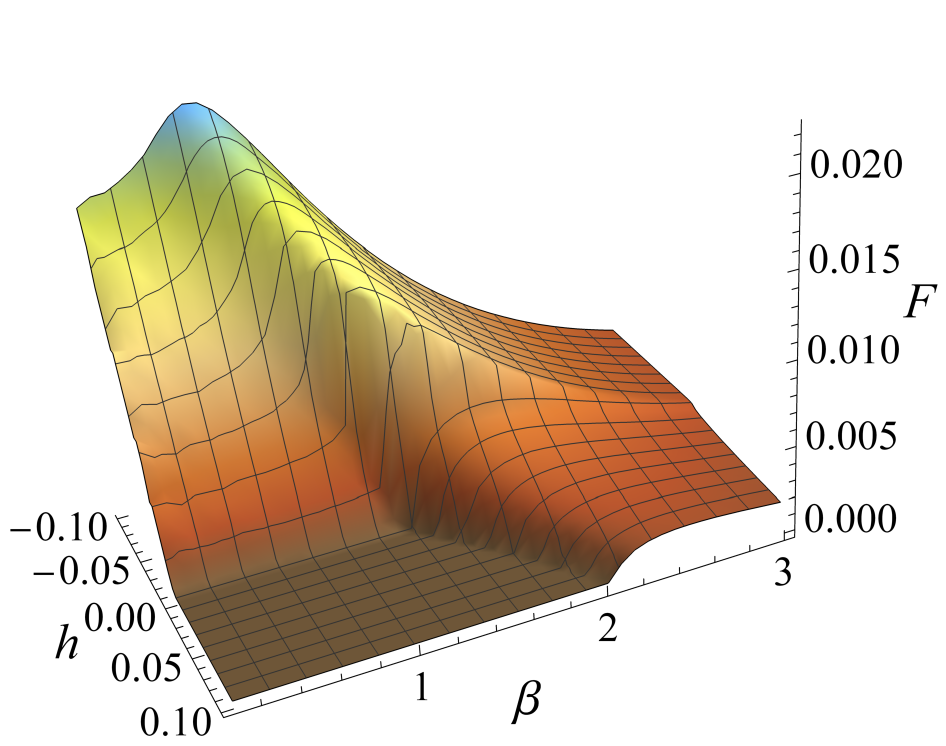}
\caption{Transition rates $F_\pm$ in the polymer quantum field theory as a function of the energy-level spacing $|h|$ and the rapidity $\beta$. }\label{fig1}
\end{figure}

Substituting the detector's trajectory in \eqref{trajectory} into the above Wightman function in \eqref{WF}, we can straightly calculate  the transition rates in \eqref{gamma} of the inertially moving detector that is coupled to a scalar field in polymer quantum field theory,
\begin{eqnarray}\label{TRRs1}
\nonumber	\gamma_\pm&=&\frac{\mu^2M_\star}{2\pi\text{sinh}\beta}\sum_{n=0}^{\infty}\int_{0}^{\infty}dgg|\sqrt{M_\star}c_{n}(g)|^2
\\
&&\times \text{H}(g\sinh\beta-|h+\omega_n(g)\cosh\beta|),
\end{eqnarray}
where $\omega_{n}(g)=\Delta E_{n}(g)/M_\star$. Since only nonzero matrix elements are $c_{4n+3}$ (for $n=0, 1, 2, \cdots$) \cite{Mortuza}, the above transition rates \eqref{TRRs1} can be rewritten as 
\begin{eqnarray}\label{TRRs2}
\nonumber	\gamma_\pm&=&\frac{\mu^2M_\star}{2\pi\text{sinh}\beta}\sum_{n=0}^{\infty}\int_{0}^{\infty}dgg|\sqrt{M_\star}c_{4n+3}(g)|^2
\\
&&\times \text{H}(g\sinh\beta-|h+\omega_{4n+3}(g)\cosh\beta|).
\end{eqnarray}
If the polymer quantization is real, the deexcitation rate of the detector may depend on the polymer energy scale $M_\star$ (may be identified as \emph{Planck energy})
and the detector's rapidity, $\beta$. Which never happens for the usual massless quantum scaler field theory. Furthermore, it has been demonstrated in Refs. \cite{Husain, Nirmalya1, Jorma, Nirmalya2} that in the polymer quantum field theory even an inertially moving detector with a constant velocity can still get excited when its rapidity exceeds a critical one. This comes from the fact that the dispersion in the polymer quantum field theory, $\Delta E_{3}$ in \eqref{TRRs2}, may satisfy the excitation condition of an inertial 
detector discussed in Section \ref{subsectionA}. Numerical calculations show that in this case the critical velocity is found to be $\beta_c\approx1.3675$. To better understand 
the excitation and deexcitation rates of the inertial detector in the polymer quantum field theory, in Fig. \ref{fig1} we plot them as a function of the detector's energy-level spacing $|h|$ and detector's rapidity $\beta$. It is found that there is a critical velocity only beyond which the detector could get excited. However, the deexcitation rate 
always exists and depends on the detector's velocity.

Since the detector's transition rates presence quite different behavior when below and beyond the critical velocity $\beta_c$ for arbitrary small $h$, in what follows 
we will separately investigate the GP acquired by the detector for two cases: 1) $\beta<\beta_c$; 2) $\beta>\beta_c$.

\begin{figure*}
\centering
\includegraphics[width=0.9\textwidth]{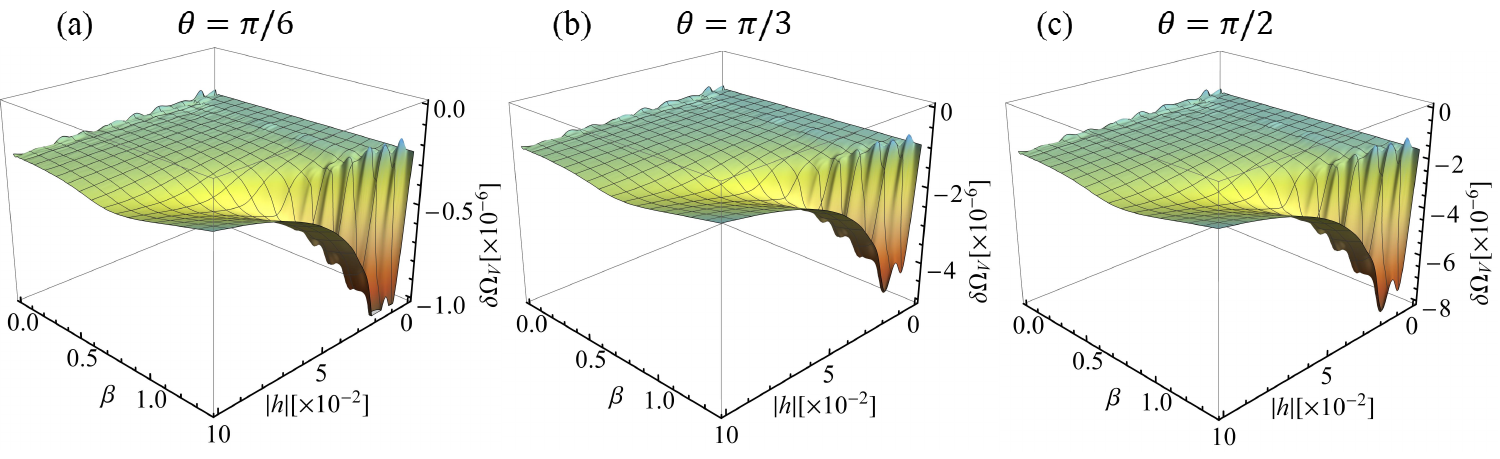}
\caption{The geometric phase as a function of the detector's energy-level spacing $|h|$ and its velocity $\beta$ for different initial states $\theta=(\pi/6, \pi/3, \pi/2)$. Here the velocity $\beta<\beta_c\approx1.3675$, the evolution time $T=2\pi/\omega_0$, the coupling strength $\mu\sim10^{-3}$ have been taken.}
\label{fig2}
\end{figure*}

\subsection{Case1: $\beta<\beta_c$}\label{section-case1}

As discussed above, when the detector's velocity is below the critical one $\beta_c$, the detector then does not get excited, and correspondingly $c_+$ in \eqref{cpm} vanishes. Therefore, in this case only $c_-$ related to the deexcitation rate has the contribution to the GP acquired by the detector.
Note that the accumulation time, seen from Eq. \eqref{correction}, can only change the magnitude of the GP correction from the interaction 
between detector and field, while does not change its behavior. Therefore, here we choose the evolution time $T=2\pi/\omega_0$ just for an example, to explore 
how the detector's energy-level spacing $|h|$ and its velocity $\beta$ affect the GP. 

Fixing a typical coupling constant $\mu\sim10^{-3}$, in Fig. \ref{fig2} the GP correction (shown in Eq. \eqref{correction}) of the detector as a function of the detector's energy-level spacing $|h|$ and its velocity $\beta$ (below the critical one $\beta_c$) is plotted with different initial state cases, $\theta=(\pi/6, \pi/3, \pi/2)$ (see \eqref{ISP} when $\tau=0$). It is found that for different initial states the GP corrections share similar behavior, while have different magnitudes. One can also find this 
result from Eq. \eqref{correction}, actually $\delta\Omega_V=\varphi^2\frac{\gamma_0}{8\omega_0}\sin^2\theta(\cos\theta-2)c_-$ in this case. Furthermore, near the critical velocity, it is shown that the GP is quite sensitive to the polymer quantization for smaller $h$. 
This is because that as shown in Eq. \eqref{cpm} $c_-=-2\pi F_-/h$ is proportional to $1/h$, and the transition rate $F_-$ behaves sharply as a function of
the velocity $\beta$ in the small $h$ and critical velocity nearby region shown in Fig. \ref{fig1}.
Since $|h|=\omega_0/M_\star$, the above result alternatively means that for fixed $\omega_0$ the GP may be sensitive to larger possible $M_\star$. 
Actually, in the polymer quantum field theory, the polymer energy scale $M_\star$ may be identified as the Planck energy, that is thought to be huge in theory. 
In this sense, the GP seems to be quite sensitive to the polymer quantization theory.

\begin{figure*}
\centering
\includegraphics[width=0.9\textwidth]{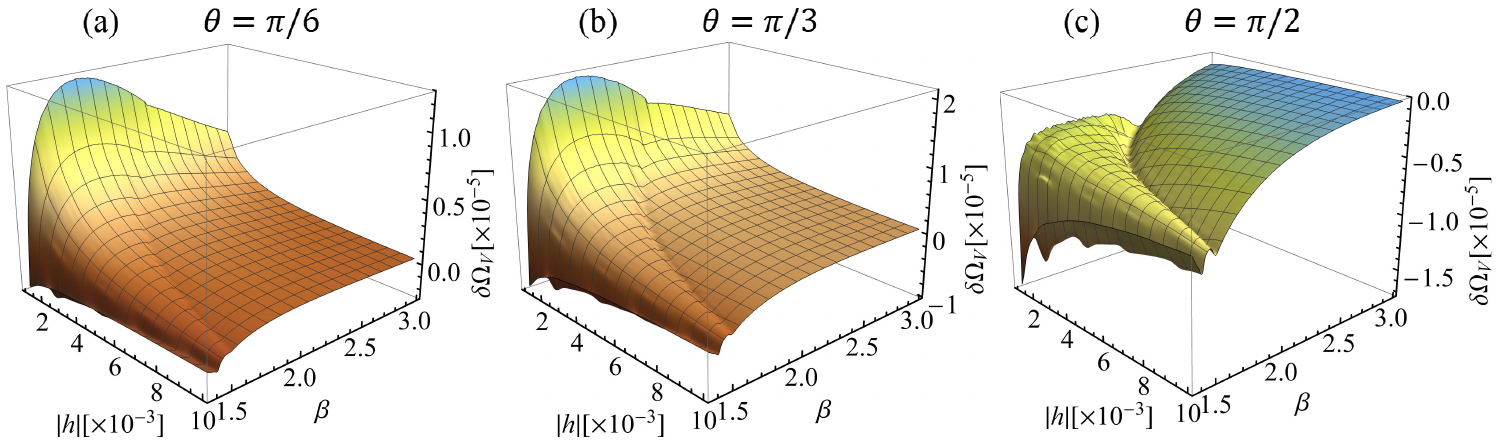}
\caption{The geometric phase as a function of the detector's energy-level spacing $|h|$ and its velocity $\beta$ for different initial states $\theta=(\pi/6, \pi/3, \pi/2)$. Here the velocity $\beta>\beta_c\approx1.3675$, the evolution time $T=2\pi/\omega_0$, the coupling strength $\mu\sim10^{-3}$ have been taken.}
\label{fig3}
\end{figure*}

\subsection{Case2: $\beta>\beta_c$}
If the detector's velocity exceeds the critical value $\beta_c$, as discussed above the inertial detector could get excited for arbitrary small energy-level spacing 
$\omega_0$ case, and correspondingly the excitation rate $c_+$ exists. Therefore, in this case both the excitation rate and the deexcitation rate have 
the contribution to the GP shown in \eqref{correction}. As a result of that, one can expect that compared with the $\beta<\beta_c$ case, the GP for the 
$\beta>\beta_c$ case may present different behavior as a function of  the detector's energy-level spacing $|h|$ and its velocity $\beta$.

In Fig. \ref{fig3}, we also take the evolution time $T=2\pi/\omega_0$ and the coupling strength $\mu\sim10^{-3}$, and plot the GP correction (shown in Eq. \eqref{correction}) of the detector as a function of the detector's energy-level spacing $|h|$ and its velocity $\beta$ (above the critical one $\beta_c$) for different initial state cases, $\theta=(\pi/6, \pi/3, \pi/2)$ (see \eqref{ISP} when $\tau=0$). We can find that the GP exhibits quite different behavior compared to the $\beta<\beta_c$ case shown in Fig. \ref{fig2}.
The maximum GP correction is acquired in arbitrary small $h$ region, while the corresponding optimal velocity (determinate the maximum GP correction)
is the critical one or others, which depends on the initial state of the detector. Note that this is because the excitation and deexcitation rates have unsymmetry responses
to the detector's energy-level spacing $|h|$ and its velocity $\beta$ shown in Fig. \ref{fig1}.
Furthermore, for high velocity and large energy-level spacing, the GP correction approaches to vanishment. It means it is quite difficult to detect the polymer quantization theory using the GP method in this case. Therefore, when the velocity exceeds the critical one, to find the maximum GP one needs to consider the initial state of the detector, and properly design the detector with the energy-level spacing $|h|$.

\begin{figure}
	\centering
	\includegraphics[width=0.38\textwidth]{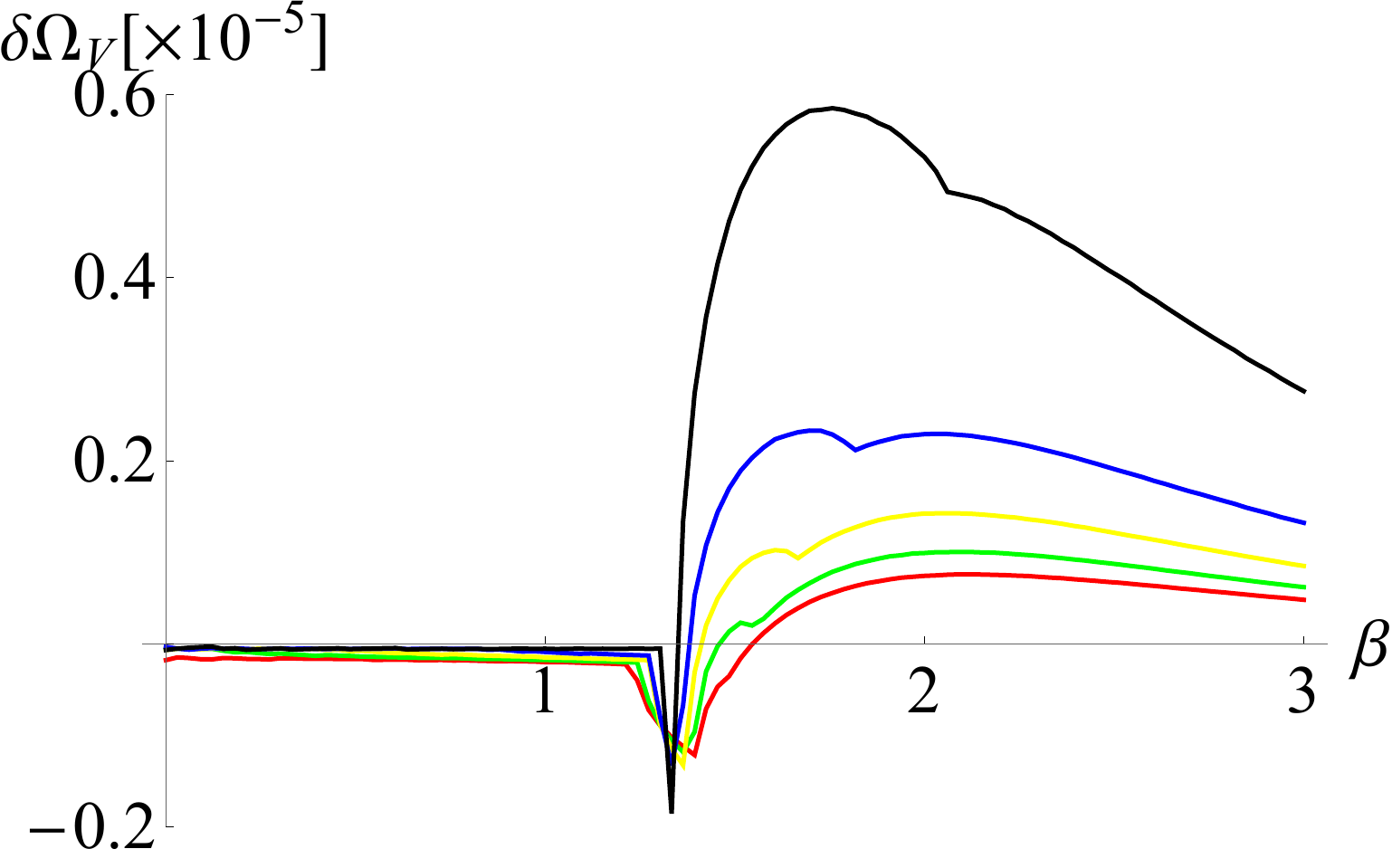}
	\caption{The geometric phase as a function of the detector's velocity $\beta$ for initial state $\theta\approx0.569\pi$. Here $|h|=(0.01, 0.008, 0.006, 0.004, 0.002)$ corresponding to the red, green, yellow, blue and black curves, respectively. The evolution time $T=2\pi/\omega_0$, the coupling strength $\mu\sim10^{-3}$ have been taken.}
	\label{fig4}
\end{figure}

In order to comparatively study how the GP is influenced in the whole velocity regime (both below and above the critical one), in Fig. \ref{fig4} we plot this GP as a function of the detector's velocity $\beta$ for initial state $\theta\approx0.569\pi$ with different fixed energy-level spacing $|h|$.
The evolution time $T=2\pi/\omega_0$, the coupling strength $\mu\sim10^{-3}$ have been taken. Here we can find that if velocity is much below the critical one the GP is insensitive to the velocity, and is much smaller than that in other velocity regime. In this case, we can obtain an approximate GP as, $\delta\Omega_V=\varphi^2\frac{\gamma_0}{8\omega_0}\sin^2\theta(\cos\theta-2)[1+2\cosh\beta\sinh^2\beta h+\mathcal{O}(h^2)]$ for small $|h|$. Near the critical velocity, $\beta_c\approx1.3675$, 
the GP shows a drastic behavior---increasing sharply as a function of the velocity. This means the GP is quite sensitive to the 
polymer quantization near the critical velocity. This has been discussed in Section \ref{section-case1}.
However, when the velocity exceeds the critical one, $c_+=\frac{2\pi F_+}{h}$ presents a drastic behavior as a function of the velocity $\beta$, and 
in this case both $c_+$ and $c_-$ have the contribution to the GP. Therefore, the GP behaves sharply near the critical velocity. Besides, it is found that 
when the velocity is above the critical one, the amount of the GP is much larger than that for the quite small velocity case (below the critical velocity). Which 
is more distinct for smaller $|h|$ case. This is because that when above the critical velocity, although $|h|$ is small, both $F_\pm$ have finite values shown in
Fig. \ref{fig1} and $c_\pm=\pm\frac{2\pi F_\pm}{h}$ are proportional to $1/|h|$. This means when above the critical velocity the GP may be susceptible for smaller $|h|$ case. Alternatively, with fixed $\omega_0$ the GP might be quite sensitive to the polymer quantization due to $|h|=\omega_0/M_\star$.

\subsection{The accumulation of GP}
\begin{figure}
	\centering
	\includegraphics[width=0.38\textwidth]{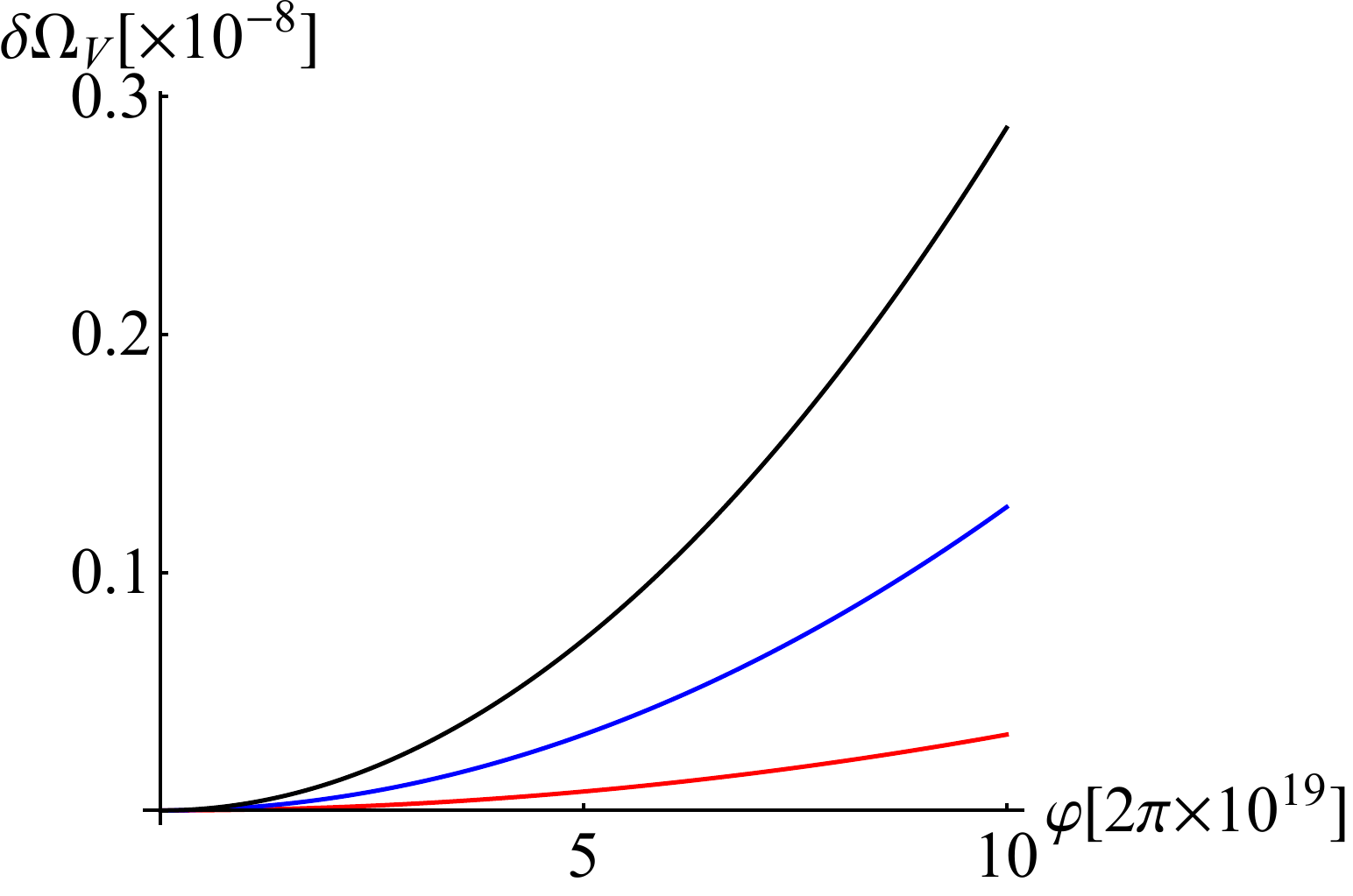}
	\caption{The geometric phase correction acquired by moving detector compared with that for the static one as a function of the detector's effective accumulation time $\varphi=\omega_0T$. Here the initial state $\theta\approx0.569\pi$, the velocity $\beta=(10^{-8}, 2\times10^{-8}, 3\times10^{-8})$ 
corresponding to the red, blue, black curves, respectively, $|h|=10^{-28}$, and the coupling strength $\mu\sim10^{-3}$ has been taken.}
	\label{fig5}
\end{figure}

Above analysis raises an interesting question: can the inertial detector moving with a constant velocity far below the critical one acquire a measurable GP?
Seen from Eq. \eqref{correction}, the GP correction is proportional to $\varphi^2=(\omega_0T)^2$, thus the GP could be accumulative. In this regard, one can expect 
that this accumulation nature may assist the inertial detector moving with a constant velocity far below the critical one to acquire a measurable GP.
In Fig. \ref{fig5}, we choose a fixed energy-level spacing $|h|$, the fixed velocity $\beta$ below the critical one and the optimal initial state $\theta\approx0.569\pi$, and plot the GP correction shown in \eqref{correction} as a function 
of the effective accumulation time $\varphi$. We can find that with the increase of the accumulation time, the GP correction increases monotonously. 

As an example, we will choose some experimentally feasible quantities and numerically estimate the possible bound of the polymer scale $M_\star$ that could be constrained using the GP method when the velocity much below the critical one (e.g., the velocity $\beta\in[0, 10^{-3}]$). Specifically, if we take the currently  feasible quantities $\omega_0\sim10^9\text{Hz}$, $\mu\sim10^{-3}$, initial state $\theta\approx0.569\pi$, and the accumulative time $T=10^7\times2\pi/\omega_0$ in the experiment, we can find that for the possible $M_\star\in[10^{13}, 10^{25}]\text{GeV}$ regime by \colb{\cite{bound}} (which means that $|h|$ ranges from $10^{-40}$ to $10^{-28}$), the GP acquired is found to be $[10^{-32}, 10^{-20}]$, which is much smaller than the minimum magnitude of GP that could be measurable with current technology. Therefore, when the velocity is much smaller than the critical one, it seems to be impossible to detect the polymer quantization using the GP method. However, if the much longer coherent time of quantum system can be maintained for the future experiments, e.g., longer than $T=10^7\times2\pi/\omega_0$ by several orders, possible polymer quantization scale discussed above might be experimentally testable using the GP method, as shown in Fig. \ref{fig5}.

Fortunately, as discussed above the GP near the critical velocity is quite sensitive to the polymer quantization. 
We wonder whether it is possible to acquire a measurable GP in this case.
For $\beta=1.3674642$ below the crucial velocity (the more precise value of the critical one is $ \beta_c=1.367464205629544$),  we also take 
$\omega_0\sim10^9\text{Hz}$, $\mu\sim10^{-3}$, initial state $\theta\approx0.569\pi$ and the accumulative time $T=10^7\times2\pi/\omega_0$, we can find that if the experimentally detectable GP is $10^{-8}$, then the possible testable lower bound of $|h|$ is $6.97\times10^{-10}$, equivalently the polymer quantization scale $M_\star\sim10^{-6}\text{GeV}$. This quantity seems quite small compared with the possible values discussed above.
However, we need to note that the chosen velocity $\beta=1.3674642$ actually is not close enough to 
the critical value such that the transition rate occurs outside the drastic response regime of velocity. This is because that 
the smaller $|h|$ is, the more drastically the transition rate changes with respect to the velocity below the critical one. Moreover,
the drastic change point of the transition rate becomes closer to the critical velocity when $|h|$ is smaller. One can see this from Fig. \ref{fig1}.
To obtain a tighter bound on the polymer quantization scale $M_\star$ using the GP method, the chosen velocity should be as close as possible 
to the critical one, while whose precision is beyond our calculation.
When the detector's velocity exceeds the critical one, $\beta_c$, in theory arbitrary 
 $M_\star$ could be testable by appropriately choosing the energy-level spacing $\omega_0$, the accumulative time $T$, and the initial state of the detector. Note that although this critical velocity seems to be quite large, actually it is smaller than the possible velocity to which the ions could be accelerated at the Relativistic Heavy Ion Collider \cite{Husain}.

%%%%%%%%%%%%%%%%%%%%%%%%%%%%%%%%%%%%%%%%%%%%%%%%%%%%%%%%%%%
\section{Discussions and Conclusions} \label{section4}
In summary, we propose a scheme to probe the Lorentz violation in quantum field theory by using the GP of a detector that is coupled to the field.
For a massless scalar quantum field with a special general dispersion relations of the form $\omega_{|\bk|}=|\bk|f(|\bk|/M_\star)$, we obtain the analytical formula of GP correction that is acquired by a two-level atomic detector. We find that unlike the Lorentz invariant case, this GP correction in the presence of Lorentz-violating quantum field theory depends on the velocity of the detector. Therefore, this velocity-dependent nature of the GP in principle could be used as a criteria to probe the Lorentz violation in quantum field theory. In particular, if the function $f$ dips below unity somewhere and $d$ is nonvanishing anywhere, transition rates in an inertial detector are strongly Lorentz violating at arbitrary low transition energies, thus correspondingly the GP correction acquired by the detector may also present a drastic low-energy Lorentz violation. Together with the ultra-high sensitivity of GP to ultraweak fields, in this sense it is expected that GP could be used as a good measurement method to probe the Lorentz violation. Besides, we also apply this GP method to the exploration of polymer quantum field theory motivated by the loop quantum gravity. We find that the detector could acquire an experimentally detectable GP with the assist of detector's velocity.

Note that our detector model here is quite closely similar to the scenario where an atom is coupled to a quantized electromagnetic field by means of 
the dipole moment interaction \cite{Breuer, Eduardo}. Therefore, our results above should apply to atoms or ions that move with a relativistic velocity. 
Near the critical velocity, the GP correction is predicted to present a dramatic response to Lorentz violation in low-energy regime. It is interesting that 
this critical velocity $\beta_c\approx1.3675$ is below the rapidity, $\beta\approx3$, to which the ions could be accelerated at the Relativistic Heavy Ion Collider \cite{Husain}. In this sense, the GP correction resulting from polymer quantum field theory, if there be, could be feasible experimentally in principle.
What is more, we also discuss whether one can detect this Lorentz violation when the detector's velocity is far below the critical one. We find that 
the accumulation nature of the GP could assist to reduce the requirement of velocity that may induce measurable GP correction in future.

This GP method can also be used to probe the nonlocal field theory \cite{Pais}. Recently, inertial particle detector method  \cite{Alessio1} 
and optomechanical experiments \cite{Alessio2, Alessio3} have been proposed to test this theory. It is shown that spontaneous emission processes of a low-energy particle detector are very sensitive to high-energy nonlocality scales \cite{Alessio1}. Given that the GP is closely related to the detector's transition rate while is much more sensitive than the transition rate from an experimental point of view, the GP thus may be susceptible to this nonlocal field theory. Future work may also focus on the tests of the polymer quantum field theory via optomechanical experiments like in Refs. \cite{Alessio2, Alessio3}, exploring how such polymer effective field theory leads to a modified Shr\"odinger evolution \cite{Mortuza} assisted with the tomography method of quantum field \cite{Tian4}.

%%%%%%%%%%%%%%%%%%%%%%%%%%%%%%%%%%%%%%%%%%%%%%%%%%%%%%%%%
\begin{acknowledgments}
We thank Viqar Husain very much for the discussions on the physics of Ploymer quantization and some key calculations in Ref. \cite{Husain}. ZT was supported by the scientific research start-up funds of Hangzhou Normal University: 4245C50224204016.

\end{acknowledgments}

\onecolumngrid
\vspace{1.5cm}

\newpage
%==========================================================================================================
%==========================================================================================================
%==================================== Supplemental Material ====================================================
%==========================================================================================================
%==========================================================================================================
\pagebreak
\clearpage
\widetext

\end{document}